\documentstyle[amssymb,aps,multicol,epsfig,prl,floats]{revtex}

\begin{document}
\draft

\twocolumn[\hsize\textwidth\columnwidth\hsize\csname @twocolumnfalse\endcsname
\author{V. Podzorov$^{1}$, M. Uehara$^{1}$, M. E. Gershenson$^{1}$, 
T. Y. Koo$^{2}$ and S-W. Cheong$^{1,2}$}
\address{$^{1}$Serin Physics Laboratory, Rutgers University, Piscataway, 
NJ 08854-8019 \\
$^{2}$Bell Labs, Lucent Technologies, Murray Hill, NJ 07974}
\date{\today }
\title{Giant 1/f noise in perovskite manganites: evidence of the percolation 
threshold}
\maketitle

\begin{abstract}
We discovered an unprecedented magnitude of the 1/f noise 
near the Curie temperature ($T_{c}$) in low-$T_{c}$
manganites. The scaling behavior of the 1/f noise and 
resistance provides strong evidence of the percolation nature of the ferromagnetic
transition in the polycrystalline samples. The step-like changes of the resistance 
with temperature, observed for single crystals, suggest that the size of the 
ferromagnetic domains depends on the size of crystallites.  

\end{abstract}


]

In the beginning of the colossal magnetoresistance (CMR) research on
manganites, it was realized that the ferromagnetic (FM) transition occurs
simultaneously with the metal-insulator transition \cite{helmholt,jin,nagaev}. 
The CMR was attributed to the magnetic-field-induced shift of
this transition \cite{cheonghwang}. The nature of the transition changes
drastically when the transition temperature $T_{c}$ is varied with the
chemical pressure\cite{cheonghwang}. While the transition in the high-$T_{c}$
materials bears a resemblance with the second-order transition\cite{greene},
the low-$T_{c}$ manganites demonstrate many features which are intrinsic to the
first-order transitions, including a strong thermal hysteresis of the
resistivity $\rho$ and magnetization $M$. There is a growing theoretical and
experimental evidence that the transport properties of the insulating state
above $T_{c}$ are dominated by small polarons or magnetic polarons, and that
the band-like carriers become important below $T_{c}$ \cite{millis1,roder,zhou}. 
Recently, it was suggested that these two types of
carriers coexist near $T_{c}$\cite{salamon}. In addition, the static
coexistence of the metallic FM phase and the insulating charged-ordered (CO)
phase was found to play an important role in the CMR effects, especially in
the low-$T_{c}$ materials\cite{masatomo}.

In this Letter, we report on measurements of the temperature dependence of
the 1/f noise in polycrystalline and single crystal samples of low-$T_{c}$
manganites. Our data strongly indicate that the so-called Curie temperature
in the low-$T_{c}$ materials is, in fact, a percolation transition
temperature rather than the temperature of the long-range ferromagnetic
phase transition. The scaling analysis of the 1/f noise is consistent with
the percolation model of conducting domains randomly distributed in an
insulating matrix\cite{tremblay}.

We have measured the 1/f noise in the poly- and single crystal bulk samples
of $La_{5/8-x}Pr_{x}Ca_{3/8}MnO_{3}$ with $x=0.35$\cite{comment1}. The sample
preparation is described elsewhere\cite{cheonghwang}. Typically, the
polycrystalline samples were $4\times 1\times 1$ $mm^{3}$, single crystals - 
$3\times 1\times 0.5$ $mm^{3}$. The spectral density of the 1/f noise, $S_{V}
$, and $\rho $ have been measured in the four-point configuration at the
temperature $T=4.2-300K$ for both cooling and warming. For the $S_{V}$
measurements, we used the Stanford Research 830 lock-in amplifier in the
mean average deviation mode with an equivalent noise bandwidth $1Hz$. The dc
voltage applied to the sample, $V$, \ and $S_{V}$ were recorded
simultaneously as a function of $T$ for the samples biased with a fixed dc
current ($10^{-5}-10^{-3}A$). It has been verified that the spectrum of the
noise has a power-law form $1/f$\ $^{\gamma }$ in the frequency range $%
f=1-10^{3}Hz$ with $\gamma $\ close to unity for all temperatures. All the
data discussed below were obtained in the linear regime, where the rms noise
was linear in current.

The upper panel of Fig. 1 shows the dependences $\rho (T)$ for the poly- and
single crystal samples. In accord with prior publications
\cite{cheonghwang,masatomo},
$\rho $ increases with cooling below the CO transition ($%
T_{CO}\sim 210K$ for $x>0.3$), reaches the
maximum and decreases rapidly when the system undergoes the transition into
the FM state. The FM transition temperature is strongly $x$-dependent: $T_{c}
$ increases from $35K$ for $x=0.4$ to $75K$ for $x=0.35$. At lower
temperatures, $\rho $ is almost $T$-independent, its value is anomalously
large even for the single crystal samples. The transition into the FM phase
is accompanied by the increase of the magnetization, which saturates at $%
T<T_{c}$ (see the lower panel of Fig. 1). In contrast to $\rho $, the
magnetization changes gradually at $T_{c}$ for both poly- and single crystal
samples. This smooth dependence $M(T)$, which is observed even in low
magnetic fields, is unusual for the FM transition in a homogeneous system. A
strong temperature hysteresis of $\rho $ and $M$ was observed for all the
samples discussed in this paper. 
\begin{figure}[ht]
\epsfig{file=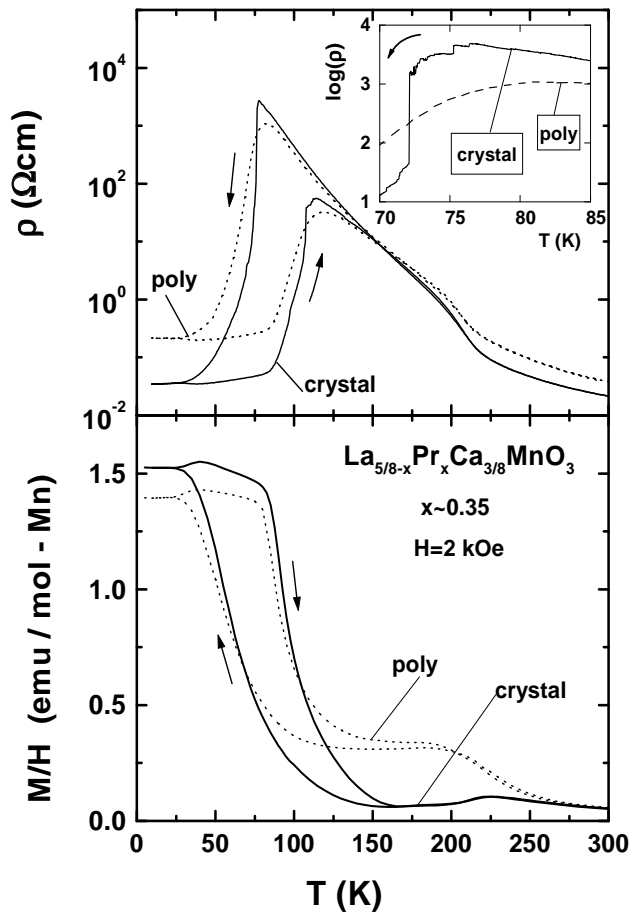, height=0.7\textwidth, width=0.5\textwidth}
\caption{The upper panel: the temperature dependences of the resistivity for
$La_{5/8-x}Pr_{x}Ca_{3/8}MnO_{3}$ ($x=0.35$) (the dashed line -
the polycrystalline sample, the solid line - the single crystal). The inset is 
the blow-up of the $\rho (T)$ dependence for cooling near $T_{c}$.
The lower panel: the temperature dependences of the
magnetic susceptibility for the same samples (the dashed line - polycrystalline, 
the solid line - single crystal) measured at $H=2 kOe$. Arrows indicate 
directions of the temperature change.}
\label{Fig.1}
\end{figure}

Qualitatively, the temperature dependence of the spectral density of the 1/f
noise is as follows: $S_{V}$ increases upon cooling below $T_{CO}$, reaches
the maximum, and decreases steeply on the metallic side of the CO-FM
transition. In the CO phase, far from the CO-FM\ transition, the increase of 
$S_{V}$ with decreasing $T$ is due mostly to the growth of $\rho$. Indeed,
in the linear regime, $S_{V}$ is proportional to $V^{2}$, or, for a fixed dc
current $I$, to $\rho ^{2}$ \cite{kogan}:

\begin{equation}
\frac{S_{V}}{V^{2}}=\frac{\alpha }{f\cdot n\cdot v_{s}}\text{ ,}
\end{equation}
where $\alpha $ is the Hooge's parameter, $f$ is the frequency at which the
noise is measured, $n$ is the concentration of the charge carriers or
''fluctuators'', and $v_{s}$ is the volume of the sample. Below we present
the 1/f noise data in the normalized form ($S_{V}/V^{2})\cdot f\cdot
v_{s}=\alpha /n.$ Figure 2 shows the temperature dependences of $\rho $ and $%
S_{V}\cdot f\cdot v_{s}/V^{2}$ for the polycrystalline sample. The
normalized magnitude of the 1/f noise is weakly $T$-dependent in the CO
phase: it varies by a factor of 2-3 over the range $T=90-150K$, though $\rho 
$ changes by 2-3 orders of magnitude over the same interval. Interestingly,
the magnitude of the 1/f noise is anomalously large even far from the
transition. (For comparison, the typical values of $\alpha /n$ are $%
10^{-21}-10^{-25}$ $cm^{3}$ for disordered metals and $10^{-18}-10^{-21}$ $%
cm^{3}$ for semiconductors \cite{kogan}).

The magnitude of the 1/f noise increases dramatically in the vicinity of the
CO-FM transition. The sharp peak of the 1/f noise enables to determine the
transition temperature with a high accuracy; below we identify $T_{c}$ with
the temperature of the maximum of $\alpha /n$. There is a correlation
between the magnitude of the noise peak and the ratio $\rho (T_{c})/\rho
(300K)$. For example, for the polycrystalline sample (Fig. 2), $\rho $
increases by 4 orders of magnitude with cooling from room temperature down
to 80 K; the normalized noise magnitude at the transition also increases by
a factor of $\sim 10^{4}$. Since the high-temperature portion of the $\rho
(T)$ \ dependences is approximately the same for all compounds with $x>0.35$%
\cite{masatomo}, the noise peak is more pronounced for materials with higher
value of $\rho (T_{c})$, e. g. with lower $T_{c}$.

It is worth mentioning that a much less pronounced increase of the 1/f noise
at the FM transition has been reported for thin films of $%
La_{2/3-x}Y_{x}Ca_{1/3}MnO_{3}$ \cite{ramirez,weissman}. The
temperature dependences of $S_{V}/V^{2}$ for the thin films differ
qualitatively from the dependences we observe for bulk samples. The authors
of \cite{ramirez,weissman} attributed the increase of the noise
below $T_{c}$ to the magnetic domains formation. 
\begin{figure}[ht]
\epsfig{file=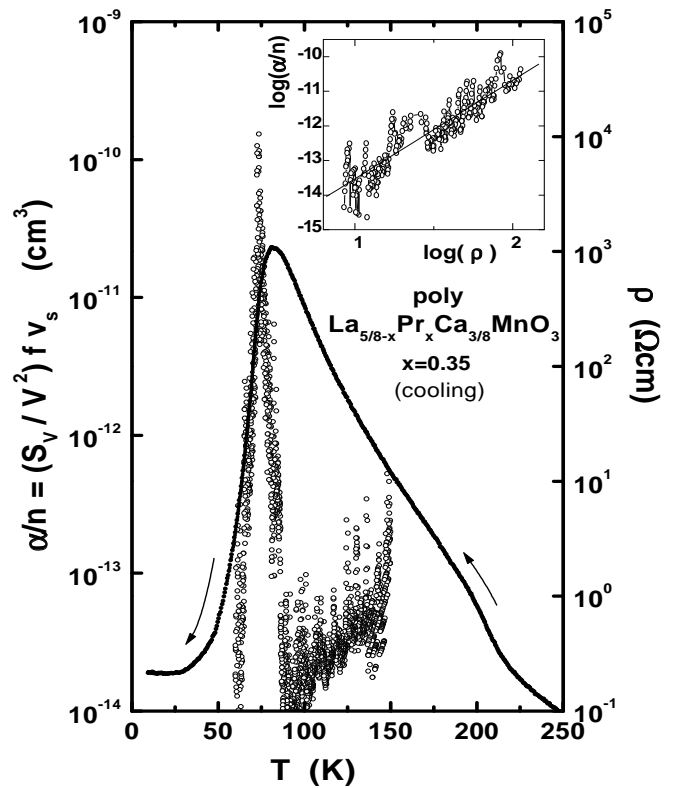, height=0.6\textwidth, width=0.5\textwidth}
\caption{The temperature dependence of the resistivity and the normalized
spectral density of the 1/f noise, $\protect\alpha /n$, for the polycrystalline
sample $La_{5/8-x}Pr_{x}Ca_{3/8}MnO_{3}$ ($x=0.35$) for cooling. The noise was
measured at $f=10Hz$; the background noise signal with no current through
the sample has been subtracted. The inset shows the scaling dependence of
the normalized magnitude of the 1/f noise versus $\rho $ in the 
interval $T=61K-73K$ (below $T_{c}=73K$). The solid line corresponds
to the power law fit $\protect\alpha /n\varpropto \rho^{2.9}.$}
\label{Fig.2}
\end{figure}
In principle, an abrupt increase of the normalized magnitude of the 1/f
noise by $1\div 2$ orders of magnitude at magnetic transitions has been
observed for many macroscopically homogeneous magnetic materials\cite
{israeloff1,israeloff2,hardner}. However, the $T$-dependence
and the magnitude of the 1/f noise in these materials differ significantly
from our data for the CMR manganites.

Our 1/f noise measurements provide strong evidence of the percolation nature
of the CO-FM transition in the polycrystalline bulk samples of the low-$T_{c}
$ manganites. These data indicate that the FM regions appear progressively
with decreasing temperature in the CO phase, and the transition occurs when
the concentration of the FM phase exceeds the percolation threshold. This is
consistent with observation of the ferromagnetic regions in manganites at $%
T>>T_{c}$, well beyond the conventional fluctuation regime (these regions
were interpreted as magnetic polarons)\cite{De Teresa}.\ A diverging
behavior of the 1/f noise is typical for the percolation metal-insulator
transition \cite{tremblay,rammal1}. At $T>>T_{c}$, the
transport properties of the sample are governed by a very large and,
apparently, weakly fluctuating contribution of the CO phase. As a result,
the 1/f noise is relatively low at $T_{c}<T<<T_{CO}$. However, the magnitude
of the 1/f noise diverges with approaching the percolation threshold. The
formation of the infinite percolation FM cluster at $T_{c}$\ is also
consistent with the observation of the maximum of $d\rho /dT$ exactly at the
same temperature. Notice that \ $T_{c}$ is lower than the temperature of the
maximum of $\rho $: this is expected for a percolating mixture of two phases
where the ''insulating'' phase has a finite $\rho $ that increases rapidly
with cooling. Previously, spatially inhomogeneous FM state and percolation
nature of the conductivity in thin films of manganites have been discussed
in Ref.\cite{babushkina}.

A clearly diverging behavior of the 1/f noise allows to determine $T_{c}$
with a high accuracy and to perform the scaling analysis of $S_{V}$ and $%
\rho $ on the ''metallic'' side of the CO-FM transition. In the vicinity of
a percolation metal-insulator transition, the scaling behavior of $\rho $
and $S_{V}/V^{2}$ is expected\cite{tremblay,kogan,rammal1}:

\begin{equation}
S_{V}/V^{2}\varpropto (p-p_{c})^{-k}\text{ ,}
\end{equation}
\begin{equation}
\rho \varpropto (p-p_{c})^{-t}\text{ .}
\end{equation}
Here $p$ is the concentration of the metallic phase, $p_{c}$ is the critical
concentration, $k$ and $t$ are the critical exponents of the noise and the
resistivity. It is convenient to represent $S_{V}/V^{2}$ as a function of $%
\rho $ (in this case, no assumption on the value of $p_{c}$ is necessary)$:$

\begin{equation}
S_{V}/V^{2}\varpropto \rho ^{-k/t}.
\end{equation}
The normalized magnitude of the 1/f noise versus $\rho $ for the
polycrystalline sample is shown in the double-log scale in the inset of Fig.
2. Within the experimental accuracy, this dependence can be fitted by the
power law (4) with $k/t=2.9\pm 0.5$. These values of $k/t$ are close to the
result $k/t=2.4$ obtained theoretically for the continuum percolation model
of conducting regions, randomly placed in an insulating matrix (the
so-called inverted random-void model)\cite{tremblay}. Previously, a similar
value of $k/t=3$ has been observed experimentally for the mixed powders of
conducting and insulating materials \cite{rudman}. Notice that the ratio of
critical exponents $k/t$ in the continuum percolation exceeds significantly $%
k/t=0.5-0.8$ for the discrete random models, though the values of the
critical exponent $t$ are almost the same for these models ($t=1.9\pm 0.03$%
). 
\begin{figure}[ht]
\epsfig{file=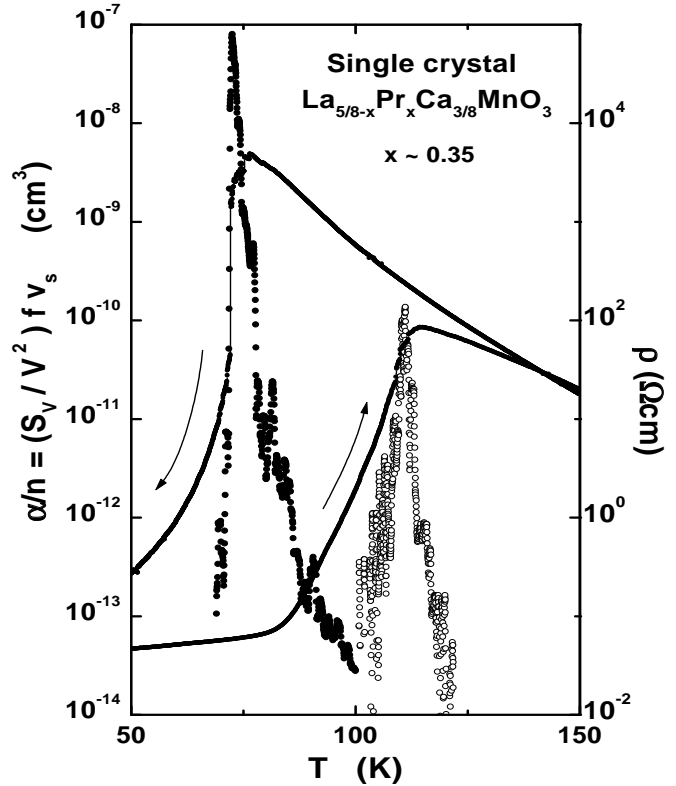, height=0.6\textwidth, width=0.5\textwidth}
\caption{The temperature dependence of the resistivity and the normalized
spectral density of the 1/f noise for cooling ($\bullet $) and warming ($%
\circ $) for the single crystal $La_{5/8-x}Pr_{x}Ca_{3/8}MnO_{3}$
($x\sim 0.35$).}
\label{Fig.3}
\end{figure}

In order to disentangle $k$ and $t$, one has to measure either the
concentration of the FM phase $p(T)$ as a function of $T$, or some quantity
that is proportional to $p$, e. g. the magnetization. In the vicinity of the
transition, $M$ is approximately a linear function of $(T_{c}-T)$ (Fig. 1).
Hence, instead of $(p-p_{c})$, we can use $(T_{c}-T)$\ as a variable in the
scaling dependences (2) and (3). The dependence $\rho (T)$ was found to be
close to the power law $\rho \varpropto (T_{c}-T)^{-t}$ with $t=2\pm 0.3$ on
the metallic side of the transition. Previously, similar values $t=2.3\pm 0.4$
and $k=5\pm 1$ has been measured for the conducting particles in an insulating
matrix \cite{chen,Lee}. The experimental values of the critical exponents $%
t=2\pm 0.3$ and $k=5.9\pm 1.5$ for our samples are consistent with the
predictions of the inverted random-void model of the continuum percolation.

For high-quality single crystals of $La_{5/8-x}Pr_{x}Ca_{3/8}MnO_{3}$ ($%
x\sim 0.35$) \cite{comment1}, we have also observed a dramatic increase of
the 1/f noise at the transition: $\alpha /n$ reaches $10^{-7}$ $cm^{3}$ for
cooling and $10^{-10}$ $cm^{3}$ for warming (Fig. 3). An evidence of the
phase inhomogeneity of the single crystal is provided by the smooth
dependence of $M(T)$ (Fig. 1), which is similar to that for the
polycrystalline samples. However, there are several important distinctions
between the temperature dependences of $\rho $ and $S_{V}/V^{2}$ for poly-
and single crystals. Although the magnitude of $\rho $ is similar for both
types of samples (see Fig. 1), $\rho $ for the single crystal exhibits
reproducible steps as a function of $T$\ in the vicinity of $T_{c}$ (see the
inset of Fig. 1). The sharp drop of $\rho $ by more than an order of
magnitude, observed for this single crystal at $T\sim 72K$, can be
interpreted as formation of a chain of a few connected FM domains between
the voltage leads. The step-like behavior of $\rho $ at the transition
indicates that the size of the FM regions in the single crystal is
significantly bigger than that in the polycrystalline samples. When the
voltage leads become ''shortened'' by a chain of the metallic FM domains, an
abrupt drop of the 1/f noise magnitude occurs. The percolation approach is
not applicable in this case, since we probe the inhomogeneous system at the
scale smaller than the percolation correlation length.

To summarize, we observed the dramatic peak of the 1/f noise in the CMR
manganite $La_{5/8-x}Pr_{x}Ca_{3/8}MnO_{3}$ ($x=0.35$ and $0.375$) at the
transition between the charge-ordered and ferromagnetic states. The scaling
analysis of the magnitude of the noise and the resistivity in the
polycrystalline samples is consistent with the continuum percolation model
of conducting FM domains randomly placed in the insulating CO matrix. Being
combined with the data on the temperature dependence of $\rho $ and $M$,
these measurements provide strong evidence of the percolating nature of the
CO-FM transition in the polycrystalline samples of the low-$T_{c}$
manganites. To the best of our knowledge, the value of $\alpha /n=S_{V}\cdot
f\cdot v_{s}/V^{2}$ $\sim 10^{-10}\div 10^{-7}cm^{3}$ observed in the low-$%
T_{c}$ manganites at the transition, is the largest normalized magnitude of
the 1/f noise for the condensed matter systems. A well-pronounced step-like
temperature dependence of the resistivity, observed for high-quality single
crystals, suggests that the scale of the phase separation is much greater
than that in the polycrystalline samples. This might indicate that the
surface energy and/or the strain effects associated with the grain boundaries
influence the size of the ferromagnetic domains in the low-$T_{c}$ manganites.

We thank Sh. Kogan for helpful discussions. This work was supported in part
by the NSF grant No. DMR-9802513.


\begin{references}
\bibitem{comment1}  Qualitatively similar data were obtained for
polycrystalline samples with $x=0.375$. The $Pr$ concentration for the
single crystal, estimated from the $T_{c}(x)$ dependence, was close to 0.35
(the nominal concentration $x=0.42$).
\end{references}
\end{document}